\documentstyle[epsf,fleqn,aps]{revtex}

\def\Jrnl#1#2#3#4{{#1} {\bf #2}, #3 (#4)}

\def\PRB{Phys. Rev. B}

\def\PRL{Phys. Rev. Lett.}

\def\Section#1{}

\def\beq{\begin{equation}}

\def\eeq{\end{equation}}

\def\bea{\begin{eqnarray}}

\def\eea{\end{eqnarray}}

\def\age{\,\raise2pt\hbox{$\mathop{>}\limits_{\raise 2pt

\hbox{$\sim$}}$}\,}

\def\ale{\,\raise2pt\hbox{$\mathop{<}\limits_{\raise 2pt

\hbox{$\sim$}}$}\,}

\setbox122 = \hbox{1}

\def\id{\rlap{1}\rlap{\kern 1pt \vbox{\hrule width 4pt depth 0 pt}}

        \rlap{\kern 4 pt \hbox{\vrule height \ht122 depth 0 pt}}

           \hskip\wd122}

\begin{document}
\tolerance 50000
\twocolumn[\hsize\textwidth\columnwidth\hsize\csname
@twocolumnfalse\endcsname

\title{Random bond $XXZ$ chains with modulated couplings}
\author{D.C.\ Cabra$^{1,2}$, A.\ De Martino$^{3}$, M.D.\ Grynberg$^{1}$,
        S.\ Peysson$^{3}$ and P.\ Pujol$^{3}$}

\address{$^{1}$Departamento de F\'{\i}sica, Universidad Nacional de la Plata,
               C.C.\ 67, (1900) La Plata, Argentina.\\
$^{2}$Facultad de Ingenier\'\i a, Universidad Nacional de Lomas de
      Zamora, Cno. de Cintura y Juan XXIII,\\
      (1832) Lomas de Zamora, Argentina.\\
$^{3}$Laboratoire de Physique,
      Groupe de Physique Th\'eorique 
      ENS Lyon, 46 All\'ee d'Italie, 69364 Lyon C\'edex 07, France.\\}

\maketitle

\vspace{.5cm}

\begin{abstract}
The magnetization behavior of $q$-periodic antiferromagnetic spin
$1/2$  Heisenberg chains under uniform magnetic fields is
investigated in a background of disorder exchange distributions.
By means of both real space decimation procedures and  numerical
diagonalizations in $XX$ chains,  it is found that for binary
disorder the magnetization exhibits wide plateaux at values of
$1+2(p-1)/q$, where $p$ is the disorder strength. In contrast, no
spin gaps are observed in the presence of continuous exchange
distributions. We also study the magnetic susceptibility at low
magnetic fields.

\vskip 0.5cm


\end{abstract}


\vskip -0.2cm
\vskip2pc]

The study of low dimensional antiferromagnets has received much
renewed attention largely owing to the synthesis of ladder
materials \cite{review}. One particular issue that captured
both experimental and theoretical efforts  in the past few years
is the appearance of magnetization plateaux, i.e. massive spin
excitations in the magnetization curve.
In general, the latter are quite robust and for pure spin systems
appear at rational magnetization values \cite{OYA,nos,Totsuka}.
More recently, some experiments have indeed confirmed these
theoretical predictions in a few particular cases \cite{S1dim} but
some issues yet remain unresolved \cite{exp}.

In order to make closer contact with experiments, one has to take
into account the disorder that is almost inevitably present due to
lattice imperfections and magnetic doping. A
relevant question related to the appearance of magnetization
plateaux is whether they are robust in the presence of
quenched disorder. As a first step in this direction, in this work
we analyze the effect of a disordered distribution of exchange
couplings with a periodically modulated mean on the magnetic
behavior of an $XXZ$ antiferromagnetic chain.
Both even \cite{Boucher} and odd
\cite{Hida} modulations are known to exist and are ultimately
responsible for the structure of the magnetization curve
\cite{OYA,nos,CG}. $q$-merized $XX$ chains have also been studied
in \cite{Zvyagin} using a Jordan-Wigner transformation.

By means of a decimation procedure similar to that used in
\cite{DM,FXXZ} we argue that plateaux in the magnetization curve
appear at specific magnetization values $m$ which depend both on
the couplings periodicity $q$ and the strength of the disorder
$p$. Hence disorder, instead of removing completely the plateau
structure, shifts the position of certain plateaux in a precise way which
depends on the disorder strength. Surprisingly,  as we shall see from our
numerical evidence, the plateaux predicted via this simple
argument are indeed present. Moreover, they are rather wide and
therefore could be eventually  observable in low temperature
experiments which in turn would allow for a precise determination
of the disorder degree.

By extending the methods of \cite{EggRie}, we also investigate the
characteristics of the magnetic susceptibility at low fields. Its
behavior shows an interesting even-odd effect, similar to that
found in the study of disordered $XX$ $N$-leg ladders
\cite{Mudry}. In fact, for $q$ odd we find the same kind of
divergence as for the homogeneously disordered case
($q=1$) \cite{Mudry}, namely
\beq
\label{chi-odd}
\chi_z \propto \frac{1}{H[\ln(H^2)]^3}\,,
\eeq
whereas the even $q$ modulations yield a generic non-universal
power law behavior as in \cite{rusos,HG} for $q=2$,
\beq
\label{chi-even}
\chi_z \propto  H^{\alpha
-1}\, ,
\eeq
where we give an analytic expression for $\alpha$ for arbitrary $q$.
In principle, these results should
emerge from experimental susceptibility measurements in disordered
low dimensional compounds.

In what follows we focus attention on the occurrence conditions of
zero temperature plateaux in a random $q$-merized $XXZ$ spin-1/2
chain whose Hamiltonian is
\beq
\label{Hamiltonian}
H = \sum_i
J_i\, \left( S^x_i S^x_{i+1} + S^y_i S^y_{i+1} + \Delta S^z_i
S^z_{i+1}\,\right)\,,
\eeq
where $J_i$ are randomly distributed
bonds. Specifically, let us consider a binary distribution of
strength $p$ ($p=0$ corresponds to the pure $q$-merized case
while $p=1$ corresponds to the uniform chain),
\beq \hspace{-0.7cm}
\label{q-bin}
P(J_i)
=  p \: \delta(J_i-J') + (1-p)\: \delta(J_i-J_0 - \gamma_i J)\,,
\eeq
where $\gamma_i \equiv \gamma,\: (-\gamma)\,$ if $i = q n,\:$ ($ i
\ne q n\,$), along with a Gaussian disorder $P(J_i) \propto \exp
-\frac{(J_i-\overline{J_i})^2} {2\,\sigma^2_i}$, and log-normal
distribution given in terms of $W_i = \ln(J_i)$ and $P(W_i)
\propto \exp -\frac{(W_i-\overline{W_i})^2} {2\,\lambda^2_i}$. All
these distributions, taken with same mean and variance, are built
to enforce $q$-merization, whose value is measured on average by
$\gamma J$. In what follows we assume that $J'$ is the smallest 
coupling and consider $ 0 < \gamma  < J_0/J $.

a. {\it Decimation procedure} --- Here we follow the arguments
used by Fisher in \cite{FXXZ}.
The procedure is roughly to decimate the strongest bonds up to an energy
scale given by the temperature. The remaining spins can be considered as free
and each of them will then give rise to a Curie behavior in
the magnetic susceptibility.

In our problem (which is at $T=0$) the energy scale is provided by
the magnetic field, and in order to compute the magnetization,
decimation has to be stopped at an energy scale of the order of
the magnetic field. We assume that all spins coupled by bonds
stronger than the magnetic field form singlets and do not
contribute to the magnetization, whereas spins coupled by weaker
bonds are completely polarized. The magnetization is thus
proportional to the fraction of remaining spins at the step where
we stop decimation. Our argument happens to apply well to the
binary distribution, provided the energy scales of the involved
exchanges are well separated.

b. {\it Plateaux in $q$-merized chains} ---Let us first consider
the case $q=2$ and assume we start at high
enough magnetic field, such that all spins are polarized
(saturation, $m=1$) and begin decreasing the magnetic field. The
magnetization stays constant for a while, then decreases abruptly
at $h \sim J_0+\gamma J$ and after that a plateau occurs at $m=p$.
This can be easily understood: at $h \sim J_0+\gamma J$ we
can decimate all the strongest bonds $J_0+\gamma J$ (the
corresponding spin pairs form singlets and do not contribute),
and the number of remaining (completely polarized) spins is
$N - 2 \times (1-p) N/2 = p N\,$. Here, the factor of 2
comes from the removal of two spins each time we decimate a bond.
Hence, the first plateau  occurs at $m=p$.
The appearance of this spin gap is due
to the fact that the remaining strongest bonds have values
$J_0-\gamma J$ \cite{values}, and all spins left from the first step of
decimation remain polarized (and the magnetization constant), until
the magnetic field decreases to $h \sim J_0-\gamma J$. At this
point the magnetization again decreases abruptly and a second
plateau occurs. The abrupt change corresponds to the decimation of
bonds $J_0-\gamma J$ which leaves us with
$ N - 2 \times (1-p)\frac{N}{2} - 2
\times (1-p) \frac{N}{2} p^2$ completely polarized bonds.
The plateau occurs then at magnetization $m = p - p^2 + p^3$.
The term $(1-p) \frac{N}{2} p^2$ comes from the bonds $J_0-\gamma
J$ which, having a $J'$ bond at each side, were not decimated in
the first step and thus is the number of bonds actually decimated
at the second step.
Evidently, for $q >2$ we can follow the same reasoning. Thus, the
number of spins which yield finite contributions to the
total magnetization at $h \sim J_0 + \gamma J$ is simply $N - 2
\times (1-p) N/q$. Hence, we find the first plateau at
\beq
m = 1 + \frac{2}{q} \,(p-1)\,.
\label{q}
\eeq
Notice that this
result locates correctly the spin gaps appearing in a pure
$q$-merized chain ($p=0$). In this sense, Eq.\,(\ref{q}) provides an
extension of this latter case \cite{CG} in the presence of binary
disorder. Since the decimation procedure applies for
generic $XXZ$ chains \cite{FXXZ}, we conclude that the emergence
of the plateaux predicted in (\ref{q}) is a generic feature, at
least with the anisotropy parameter $\vert \,\Delta \,\vert <1$.
It is straightforward to generalize this analysis to the case of
an arbitrary but {\it discrete} probability distribution. Given
a finite difference between the highest values of the
couplings in the inequivalent sites, one can predict the presence
and position of the plateaux.

To enable an independent check of these assertions, we turn to a
numerical diagonalization of the Hamiltonian (\ref{Hamiltonian})
contenting ourselves with the analysis of the subcase $\Delta =
0$. This allows us to explore rather long chains using a fair
number of disorder realizations. In Figs.\, 1(a), 1(b) and
1(c) we show respectively the whole magnetization curves obtained
for $q=\,$2, 3 and 4 after averaging over 100 samples of $L= 5
\times 10^4$ sites under the exchange disorder (\ref{q-bin}). It
can be readily verified that this set of robust plateaux emerges
quite precisely at the critical magnetizations given by
Eq.\,(\ref{q}). The secondary plateaux, though narrower, are still
visible in Fig.\ (1).

It is important to stress that the derivation of our results
for the quantization conditions derived above
rely strongly on the
discreteness of the probability distribution and would not be
applicable to an arbitrary continuous exchange disorder. In fact, for the
Gaussian case referred to above it turns out that  no traces of
plateaux can be observed.
Furthermore, the magnetic susceptibility in the Gaussian case
only vanishes asymptotically when approaching the saturation
regime, as can be seen in Fig. 2 for a variety of coupling
periodicities. Here, the sampling was improved up to $5 \times
10^4$ realizations though the length of the chain was reduced to
$L=1500$ sites, as the CPU time per spectrum grows as $L^2$.

>From the numerical curves, it appears that the usual DN-PT
\cite{DNPT} transition is smoothed in the presence of disorder,
both in the cases of binary and Gaussian distribution. Using the
exact results of \cite{dyson} for a family of Poissonian
distributions, one sees that the behavior of the magnetization
close to saturation has a non-universal exponential decrease. On
the other hand, this non-universality is reflected for the binary
case, in the fact that saturation occurs with an upper bound given
by $2 J_{\text{max}}$. Since the universal DN-PT transition
is destroyed near saturation, we expect that the same will occur
in the vicinity of a non-trivial plateau. In fact, this is
noticeable in the numerical data.

c. {\it Susceptibility at low magnetic fields} ---
For homogeneously disordered chains ({\it i.e.} $q=1$),
one can use the decimation procedure of
\cite{FXXZ} along with the universality of the fixed point,
to show that either for discrete or continuous distributions
the low field magnetic susceptibility behaves according to
Eq.\, (\ref{chi-odd}).
Following a simple argument based on
random walk motion used in \cite{EggRie}, it can be
readily shown that for $\Delta =0$ (or $XX$ chains),
these arguments can be extended to the case of $q$ {\it odd}
giving the same singularities.
In fact, these expectations can be compared to the numerical results
obtained for $q$ odd with both Gaussian and binary disorders,
as shown in Figs.\ 2 and 3 respectively.
In particular, we direct the reader to the semi-log insets
of Figs.\ 2(b), 2(d), 3(b) and 3(d) which evidently follow
the universal singularity referred to in Eq.\, (\ref{chi-odd}).
The numerical results for the log-normal distribution lead to
the same qualitative behavior obtained for the Gaussian case.

For $q$ {\it even}, for which there is a plateau at $m=0$ in the pure
case, the situation is more subtle. Using the notation of
\cite{EggRie}, for $XX$ chains we can again define a
random walk of the variable $u_i = \ln(\Delta_i)$ between the
boundaries $\ln(\tilde{V}^2/E)$ and $\ln(E)$,  now with a
driving force $F$ and diffusion coefficient $D$ given by
\bea
\label{param1}
\hskip -0.4cm
F &=& \frac{2}{q}  < \ln J_{i = qn}^2  - \ln J_{i \neq qn}^2 >\,,\\
\label{param2}
\hskip -0.4cm D &=& \frac{1}{q} \left[ \text{var}^2 (\ln J_{i=qn}^2) +
(q-1) \text{var}^2 (\ln J_{i\neq qn}^2) \right].
\eea
By means of the method given in \cite{EggRie}
for the undriven random walk, one can
approximate the problem to a discrete time diffusion problem with an absorbing
and reflecting wall. One can then show that the average number of bonds for a
cycle to be completed goes now like $\bar{n} \sim e^{\,\alpha
\,\Delta u /2}$,
which gives the asymptotic behavior for the magnetic
susceptibility as in (\ref{chi-even}). The non-universal
exponent $\alpha$ turns out to depend on the distribution
parameters (\ref{param1}) and (\ref{param2})
namely, $\alpha = 2 F/D$.
Also, this exponent coincides with the results obtained 
in \cite{HG} using decimation and other methods 
for the $q=2$, $XXZ$ chain.

Once more, our numerical results
for the binary and Gaussian coupling distributions considered above
lend further support to this non-universal picture of even modulations.
Specifically, there are in fact situations for which the
low field susceptibility can either diverge ($\alpha < 1$)
as displayed in Figs.\ 2(a), 2(c), 3(c) or collapse
($\alpha > 1$) as shown in Figs.\ 3(a).
Moreover, we checked that our $\alpha$ exponents
can fit reasonably the numerical behavior obtained in this regime,
as indicated by the log-log insets of Figs.\ 2(a) and 2(c).

To summarize, we have studied the effect of disorder on the
plateaux structure in $q$-merized $XXZ$ chains. By means
of a simple real space decimation procedure we could account
for a non-trivial phenomenon namely, the shift in the
magnetization values for which certain plateaux emerge, as compared to the
pure system. This was tested by numerical diagonalizations of large
XX chains finding a remarkable agreement with Eq.\ (\ref{q}).
We have also analyzed the behavior of the low magnetic field
susceptibility which exhibits a clear $q$-odd (even)
logarithmic (power law) behavior.
Our theoretical predictions could be experimentally checked on
dimerized compounds such as CuGeO$_3$ doped with Si \cite{SiCuGeO}
under magnetic fields. Also trimerized compounds exist in nature
\cite{Hida}
and it would be interesting to see if they can be doped.
We trust this work will convey an interesting motivation for
further experimental studies.

\vskip 0.15cm

We acknowledge useful discussions with P.\ Degiovanni,
F.\ Delduc, A.\ Honecker, R.\ M\'elin, and C.\ Mudry.
This work was done under partial support of EC TMR contract
FMRX-CT96-0012. The research of D.C.C and M.D.G. is
partially supported by  CONICET,
ANPCyT (grant No.\ 03-02249) and Fundaci\'on Antorchas,
Argentina (grant No.\ A-13622/1-106).


\newpage
\hbox{%
\vspace{-1.8cm}}

\begin{figure}
\hbox{%
\epsfxsize=5.1in
\hspace{-2.7cm}
\epsffile{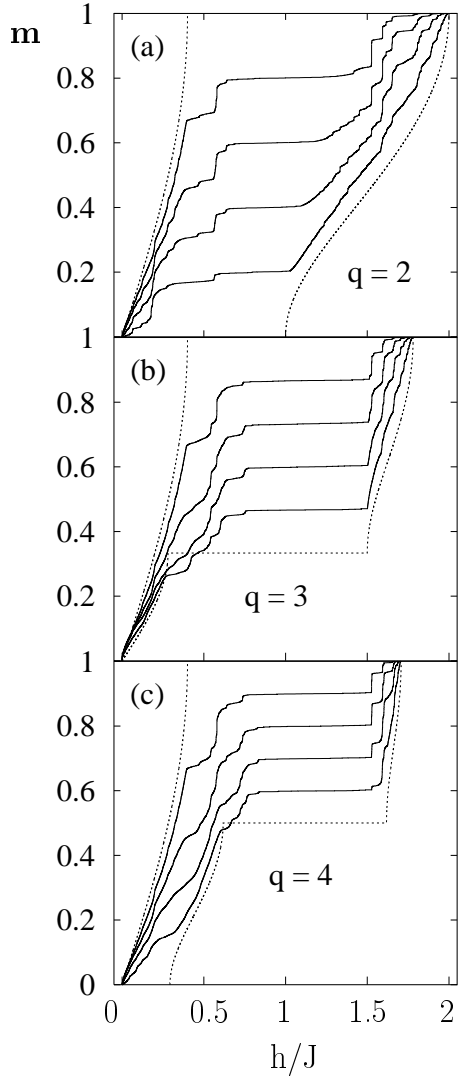}}
\vspace{ -2cm}
\caption{Magnetization curves of modulated $XX$ spin chains
with $q = 2\,$ (a), 3 (b) and 4 (c),
immersed in disordered binary backgrounds of strength $p$.
Solid lines represent averages over 100 samples with
$5 \times 10^4\,$ sites, $J'/J_0 = 0.2,\, \gamma J = 0.5\,$ and
$p = 0.2,\, 0.4,\, 0.6,\, 0.8\,$
in ascending order. The left and rightmost dotted lines
denote respectively the pure uniform and pure modulated cases
$p = 1\,$ and $p = 0\,$. \label{fig1} }
\end{figure}

\newpage
\hbox{%
\vspace{-1.8cm}}

\begin{figure}
\hbox{%
\epsfxsize=4.4in
\hspace{-1.1cm}
\epsffile{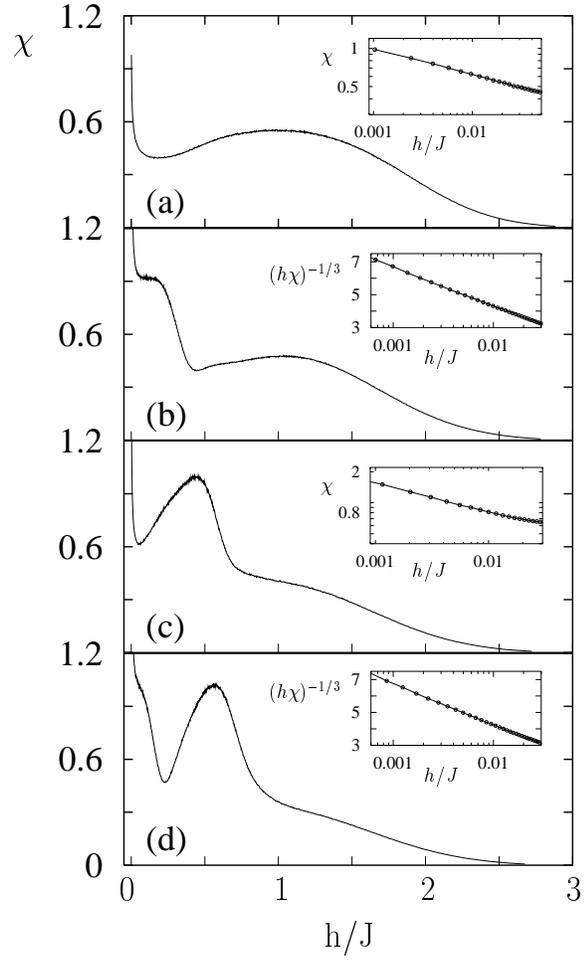}}
\caption{Magnetic susceptibility of $q$-merized
$XX$ chains after averaging over $5 \times 10^4$ samples
with 1500 sites, using  $q$-periodic Gaussian
exchange distributions for $q = 2\,$ (a), 3 (b), 4 (c),
5 (d) and strength $p=0.4$. The insets show respectively the
susceptibility behavior at low magnetic
fields which follows closely the regimes predicted
by Eqs. (1) ($q$ odd), and (2) ($q$ even),
in the text.\label{fig2} }
\end{figure}

\newpage
\hbox{%
\vspace{-1.5cm}}

\begin{figure}
\hbox{%
\epsfxsize=4.4in
\hspace{-1.1cm}
\epsffile{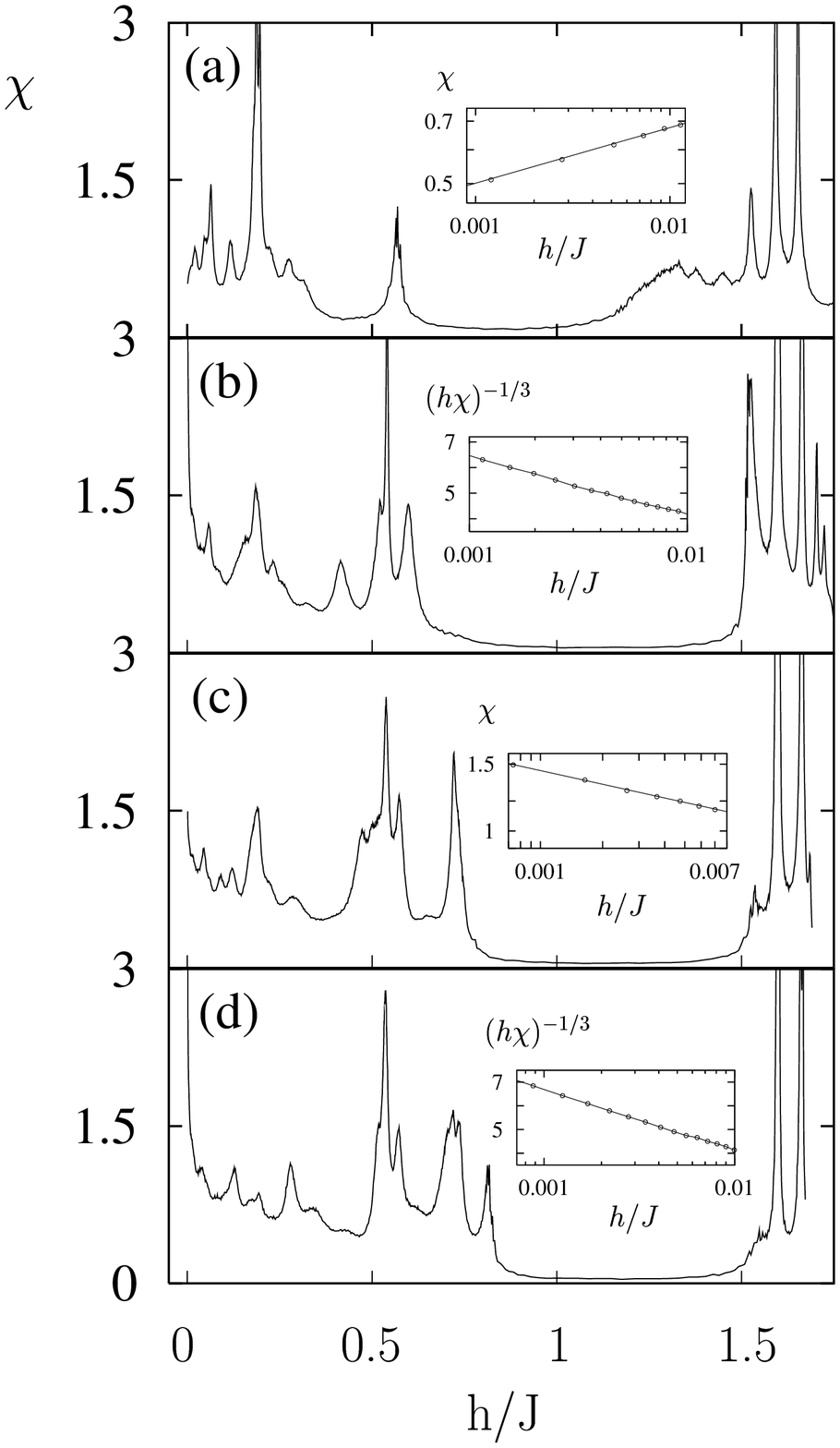}}
\vspace{-0.3cm}
\caption{Same as Fig. 2, but averaging over
a binary exchange disorder of strength $p = 0.4$.
The small field susceptibility behavior
displayed in the insets reflect the typical singularity
of odd periodicities [\,$q = 3$ (b) and 5 (d)\,],
whereas even modulated distributions [\,$q = 2$ (a) and 4 (c)\,],
are non-univesal in this regime. \label{fig3} }
\end{figure}


\begin{thebibliography}{99}
\bibitem{values} In fact, this depends on the values of
$J',\, J_0, \,\gamma J$. This is generically true for
$J_0 + \gamma J > J_0 - \gamma J > J'$ which is
satisfied by the values used in the numerical
diagonalizations.
\end{thebibliography}
\end{document}